\documentclass[aps,pra,superscriptaddress,showpacs,showkeys,a4paper,10pt,notitlepage,twocolumn]{revtex4-2}
\usepackage[utf8]{inputenc}
\usepackage[english]{babel}
\usepackage{amsmath}
\usepackage{amssymb}
\usepackage{amsfonts}
\usepackage{graphicx}
\usepackage{braket}
\usepackage{diagbox}
\usepackage{colortbl}
\usepackage{cancel}
\usepackage{upgreek}
\usepackage{mathtools}
\usepackage[T1]{fontenc}

\usepackage{xcolor}
\definecolor{link_blue}{RGB}{52,46,157}

\usepackage{longtable}
\usepackage[colorlinks,citecolor=link_blue,linkcolor=link_blue,urlcolor=link_blue]{hyperref}
\usepackage[tiny,center,uppercase]{titlesec}

\usepackage{tikz,xcolor}
\definecolor{lime}{HTML}{A6CE39}
\DeclareRobustCommand{\orcidicon}{%
 \begin{tikzpicture}
    \draw[lime, fill=lime] (0,0)
    circle [radius=0.14]
    node[white] {{\fontfamily{qag}\selectfont \tiny ID}};
    \draw[white, fill=white] (-0.0625,0.095)
    circle [radius=0.007];
    \end{tikzpicture}
    \hspace{-2mm}
}

\renewcommand{\vec}{\boldsymbol}

\begin{document}

\title{Approximating Hamiltonian for Hartree-Fock solutions for nonrelativistic atoms}

\newcommand{\orcidIDF}
{\href{https://orcid.org/0000-0003-0476-8634}{\orcidicon}}

\newcommand{\orcidODS}
{\href{https://orcid.org/0000-0002-2875-0140}{\orcidicon}}

\newcommand{\orcidSan}
{\href{https://orcid.org/0000-0003-2919-5414}{\orcidicon}}

\author{N.\ Q.\ San\orcidSan}
\email[Corresponding author: ]{nguyenquangsan@hueuni.edu.vn}
\affiliation{School of Engineering and Technology - Hue University, Hue, Vietnam}

\author{O.\ D.\ Skoromnik\orcidODS}
\affiliation{Currently without university affiliation}

\author{V.\ V.\ Triguk}
\affiliation{Currently without university affiliation}

\author{I.\ D.\ Feranchuk\orcidIDF}
\affiliation{Atomicus GmbH Amalienbadstr. 41C, 76227 Karlsruhe, Germany}
\affiliation{Belarusian State University, 4 Nezavisimosty Ave., 220030 Minsk, Belarus}

\begin{abstract}
    In our work we construct a Hamiltonian, whose eigenstates approximate the
    solutions of the self-consistent Hartree-Fock equations for nonrelativistic
    atoms and ions. Its eigenvalues are given by completely algebraic
    expressions and the eigenfunctions are defined by Coulomb wave-functions
    orbitals. Within this approximation we compute the binding energy,
    ionization potentials, electron density distribution, electron density at
    the nucleus, and atomic scattering factors of nonrelativistic atoms and
    ions.  The accuracy of our results is comparable with those obtained
    via the usage of the Hartree-Fock method but does not require solving
    integro-differential equations or numerically computing integrals with
    complex functions. This approach can serve as a good initial approximation for
    performing more accurate calculations and for the quantitative evaluation
    of physical parameters that depend on the electron density of atoms. The
    proposed approach is of interest for various fields of condensed matter
    physics, plasma physics and quantum chemistry.
\end{abstract}

\maketitle

\section{Introduction}
\label{sec:introduction}
The development of relatively simple and universal, but at the same time
sufficiently accurate and effective methods for describing the electronic
structure of atoms and ions is of great interest for numerous applications in
condensed matter physics and quantum chemistry
\cite{doi:10.1063/1.455064,*doi:10.1063/1.460447,jensen2007introduction,szabo1989modern,
RevModPhys.23.69,Skoromnik_2017,Dzikowski_2021}. One of the most well-known and
successful achievements in this field has been the Thomas-Fermi statistical
model (TFM) with various corrections \cite{Thomas_1927,*FermiA1926medoto}. This
model forms the basis for applied calculations of the structure of molecules
and solids, which are performed using the currently widely-used density
functional theory (for example, \cite{engel2011density}).  The TFM has unique
features. It does not contain any empirical parameters. It can be derived by
applying the quasi-classical approximation in solving the Hartree-Fock
equations for a nonrelativistic multi-electron system. Additionally, it is
asymptotically accurate as $Z \rightarrow \infty$ ($Z$ is the nuclear charge).
However, the model also has several well-known disadvantages. These arise from
the asymptotic nature of quasiclassics. For instance, it provides an incorrect
description of the electron density near the nucleus and in the atom's external
region. It also cannot account for shell oscillations. Moreover, it is limited
in capturing the individual properties of specific atoms when calculating their
observable characteristics.

The quantitative description of nonrelativistic many-electron atoms relies on
the numerical solution of self-consistent equations for single-electron wave
functions within the Hartree-Fock model (HFM)
\cite{HartreeA1928wave-2,*HartreeA1928wave-1,*FockA1930naeherungs}. A recent
review of studies calculating atomic properties using this method is presented
in the book by \cite{froese2022computational}. Other numerical approaches for
calculating the electronic structure of atoms exist, such as the effective
potential method \cite{grabo1997optimized} and the algorithm for the direct
numerical solution of the Schrödinger equation \cite{zhang2016efficient}.
Despite the high accuracy of numerical calculations, the construction of
approximate analytical expressions for atomic wave functions and other
characteristics remains of great interest for many applications.

The most common approach involves Slater-type orbitals
\cite{slater1932analytic} and other methods for interpolating numerical results
using analytical functions. These functions often have a relatively small
number of fitting parameters for each orbital (see \cite{zhang2016optimal}).
Analytical models of atoms, where electron wave functions take the form of
Coulomb orbitals with phenomenological parameter values like screening
constants or quantum defects, are also used (e.g.,
\cite{green1969analytic}). In \cite{feranchuk2002new}, it was shown that these
parameters can be calculated for all atoms of the periodic table using the
operator method for solving the Schrödinger equation. This calculation enabled
the construction of an accurate analytical approximation for atomic scattering
factors.

Recently, \cite{Skoromnik_2017,Dzikowski_2021} developed an analytical model of
the atom in which atomic characteristics are determined by integrals with
Coulomb orbitals. However, it requires calculation of large number of integrals
with special functions, which becomes cumbersome in higher orders of
perturbation theory. This limits the ability to analytically study the
dependence of atomic properties on the nuclear charge.

In this paper, we propose a new approach to calculate the characteristics of
nonrelativistic atoms and ions without solving differential equations or
computing integrals with numerical functions. Our method does not rely on
interpolating solutions of the Hartree-Fock equations. Instead, it involves
constructing a Hamiltonian whose eigenstates approximate these solutions. Known
as the method of approximating Hamiltonians \cite{bogolyubov1984some}, this
approach has proven effective in statistical physics.  The eigenfunctions of
the Hamiltonian constructed in our work are defined by a set of Coulomb
orbitals. The eigenvalues are determined through algebraic expressions. Using
this Hamiltonian, we can address various problems, making this approach a
viable alternative to the HFM for performing mass calculations. Our method is
particularly useful in applications like computer codes for plasma simulations,
such as CRETIN \cite{Scott2001689}, FLYCHK \cite{Chung20053}, and LASNEX-DCA
\cite{Joensson20132197}. In these cases, the TFM is used to determine the
electrostatic potential \cite{SchulzeHalberg2013323}. Our approach is also
applicable in crystallography for calculating x-ray scattering factors
\cite{Doyle:a05916}, where fits of electronic density are utilized.

As shown in the paper, the usage of the approximating Hamiltonian allows one to
calculate binding energy for atoms or ions with any nucleus charge with
relative accuracy $\delta < 1\%$ compared to the HFM results. It also enables
finding the wave function and electron density, which have correct asymptotics
at small and large distances from the nucleus and describe shell oscillations.
Additionally, it allows the calculation of ionization potential and atomic form
factors that are in good agreement with HFM results.

The paper is organized as follows: In Section \ref{sec:appr hamiltonians}, the
derivation of the approximating Hamiltonian is discussed, and its eigenvalues
and eigenfunctions are determined. Section \ref{sec:cal-char} presents the
results that can be obtained for the observable characteristics of atoms and
ions based on this Hamiltonian.

\section{Approximating Hamiltonian for nonrelativistic atom}
\label{sec:appr hamiltonians}

Let us consider the Hamiltonian of a nonrelativistic atom in the
second-quantized representation, where we use as a basis Coulomb orbitals
$|\lambda > = \psi_{\lambda}(\vec r) = R_{nl}(r) Y_{lm}(\theta,\phi) \chi_s$
with quantum numbers $\lambda = \{nlms\}$. Assume that these orbitals depend on
the effective nucleus charge $Z_{\lambda}$, which in the general case differs for
different states and does not coincide with the nucleus charge $Z$ (in our work
we utilize atomic units. The description of notations is provided in
\cite{Skoromnik_2017}):

\begin{eqnarray}
\label{1}
\hat H = - \sum_{\lambda}\frac{Z_{\lambda}^2}{2 n^2}\hat{a}^\dagger_{\lambda}\hat{a}_{\lambda} - \sum_{\lambda,\lambda'}\bra{\lambda^\prime}\frac{Z-Z_{\lambda}}{r}\ket{\lambda}\hat{a}^\dagger_{\lambda^\prime}\hat{a}_{\lambda} +\nonumber \\ \frac{1}{2} \sum_{\lambda,\lambda^\prime,\lambda_1,\lambda^{\prime}_1}\bra{\lambda_1,\lambda^{\prime}_1}\frac{1}{|\vec r_1 - \vec r_2|}\ket{\lambda,\lambda^\prime}\hat{a}^\dagger_{\lambda_1^\prime}\hat{a}^\dagger_{\lambda_1}\hat{a}_{\lambda^\prime}\hat{a}_{\lambda}
\end{eqnarray}

To extract from (\ref{1}) the approximating Hamiltonian (AH) we use the operator
method (OM) of nonperturbative description of quantum systems
\cite{OM,OM_book}. According to OM, the initial exact Hamiltonian operator is
split in such a way that in the zeroth-order approximation only those terms that commute with the number operator $\hat{n}_{\lambda} =\hat{a}^\dagger_{\lambda}\hat{a}_{\lambda}$
are kept. The remaining parts of the exact
Hamiltonian are treated as a perturbation. In our particular problem, the
operator in the zeroth-order approximation takes the following form:

\begin{eqnarray}
\label{2}
\hat H_0 = - \sum_{\lambda}\frac{Z_{\lambda}^2}{2 n^2}\hat n_{\lambda} - \sum_{\lambda }\bra{\lambda}\frac{Z-Z_{\lambda}}{r}\ket{\lambda}\hat n_{\lambda}+ \nonumber\\  \sum_{\lambda\geq\lambda^\prime}\bra{\lambda,\lambda^\prime}\frac{1}{|\vec r_1 - \vec r_2|}\ket{\lambda,\lambda^\prime}\hat n_{\lambda}\hat n_{\lambda^\prime}-\nonumber \\ \sum_{\lambda}\bra{\lambda,\lambda}\frac{1}{|\vec r_1 - \vec r_2|}\ket{\lambda,\lambda}\hat n_{\lambda}=\nonumber \\  - \sum_{\lambda}\left[\frac{Z_{\lambda}^2}{2 n^2}+ \frac{(Z-Z_{\lambda})}{n^2}Z_{\lambda}\right] \hat n_{\lambda}+\nonumber \\ \frac{1}{2}\sum_{\lambda\neq\lambda^\prime}\bra{\lambda,\lambda^\prime}\frac{1}{|\vec r_1 - \vec r_2|}\ket{\lambda,\lambda^\prime}\hat n_{\lambda}\hat n_{\lambda^\prime}.
\end{eqnarray}

The Hamiltonian (\ref{2}) includes rather cumbersome matrix elements of the
electron interaction operator. However, it can be simplified and reduced to an
algebraic form using the ansatz we derived. This approximation reduces
two-particle matrix elements to single-particle ones. To express this
relationship, let us isolate the quantum number of a shell, i.e., $
\ket{\lambda} = \ket{n,\mu} $, where $ \{\mu\}= \{lms\}$. Then, the following
approximate equality can be written:

\begin{eqnarray}
\label{3}
\int d\vec r_1 \int d\vec r_2 \psi_{n,\mu}^2(\vec r_1) \frac{1}{|\vec r_1 - \vec r_2|}\psi_{n',\mu'}^2(\vec r_2) \approx \nonumber \\  k_{n'l'}  \int d\vec r_1 \psi_{n,\mu}^2(\vec r_1)\frac{1}{r_1}  =Z_{\lambda} \frac{k_{n'l'}}{n^2};  \nonumber\\k_{nl} = 1 - \frac{\alpha}{n} - \beta \frac{l(l+1)}{n^2};\quad n' < n.
\end{eqnarray}
here $\alpha, \beta$ are empirical parameters. Their numerical values will be
determined below. The obtained relation is based on the fact that the average
radius of the shell is determined mainly by the quantum number $n$, so
$\bra{n^\prime }r_2\ket{n^\prime } \ll \bra{n}r_1\ket{n}$ and it can be
neglected with an accuracy of corrections that define the coefficient $k$.

For electrons within the same layer $n^\prime = n$, we use the following
ansatz, which also allows one to express two-particle operators through
single-particle operators
\begin{eqnarray}
\label{4}
\bra{n,\mu ; n, \mu^\prime}\frac{1}{|\vec r_1 - \vec r_2|}\ket{n,\mu;n, \mu^\prime } \approx \nonumber \\  \frac{5}{16} \bra{n,\mu;n, \mu^\prime}\frac{k^\prime_{nl}}{r_1}+\frac{k^\prime_{nl^\prime}}{r_2} \ket{n,\mu;n, \mu^\prime } = \nonumber\\ \frac{5}{16 n^2}(Z_{nl}k^\prime_{nl^\prime} + Z_{nl^\prime} k^\prime_{nl}); \nonumber \\ k^\prime_{nl} = 1+ \beta \frac{l(l+1)}{n^2}.
\end{eqnarray}

As a result, the operator (\ref{2}) is approximately represented as
\begin{eqnarray}
\label{2a}
\hat H_0 \approx \hat H_A =
- \sum_{\lambda}\left[\frac{Z_{\lambda}^2}{2 n^2}+ \frac{(Z-Z_{\lambda})}{n^2}Z_{\lambda}\right]\hat n_{\lambda}+ \nonumber \\   \sum_{\lambda>\lambda^\prime , n>n^\prime}\frac{Z_{\lambda}}{n^2}k_{\lambda^\prime}\hat n_{\lambda}\hat n_{\lambda^\prime}+ \frac{5}{16}\sum_{\lambda>\lambda^\prime, n=n^\prime}\frac{Z_{\lambda}}{n^2}k^\prime_{\lambda^\prime}\hat n_{\lambda}\hat n_{\lambda^\prime}.
\end{eqnarray}

Consequently, AH takes diagonal form and its eigenfunctions are state vectors
with a specific set of occupation numbers $g_{\lambda}$, which are $0$ or $1$.
They are determined by the eigenvalues of the particle number operators
\begin{eqnarray}
\label{5}
 \hat H_A \ket{\{g_{\lambda}\}} = E(Z,N,Z_{\lambda})\ket{\{g_{\lambda}\}};\nonumber\\
  \hat n_{\lambda} \ket{\{g_{\lambda}\}} = g_{\lambda}\ket{\{g_{\lambda}\}}; \nonumber\\
  \sum_{\lambda}g_{\lambda} = N.
\end{eqnarray}

The effective charge will be determined in accordance with the OM
\cite{OM_book}. Namely, $Z_{\lambda}$ is determined
from the fact that the eigenvalues should not depend on its
particular value for any given quantum number $\lambda$. This implies that
\begin{eqnarray}
\label{5a}
   \frac{\partial E(Z,N,Z_{\lambda})}{\partial Z_{\lambda}}= 0,
\end{eqnarray}
for any $\lambda$.

Taking this equality and (\ref{2a}) into account, we find
\begin{eqnarray}
\label{6}
\hat H_A = - \sum_{\lambda}\frac{Z_{\lambda}^2}{2 n^2}\hat n_{\lambda}; \ E(Z,N) = - \sum_{\lambda}\frac{Z_{\lambda}^2}{2 n^2} g_{\lambda}\nonumber                      \\
 Z_{\lambda} = Z - \sum_{\lambda^\prime}g_{\lambda^\prime}\kappa_{n, l^\prime} -
\frac{5}{16}\left(\sum_{\lambda''} g_{\lambda''}\kappa^\prime_{n, l''} - \kappa^\prime_{nl}\right);
\end{eqnarray}
where
\begin{eqnarray}
\label{7}
\sum_{\lambda^\prime}&=& \sum_{n^\prime = 1}^{n-1}\sum_{l^\prime = 0}^{n^\prime -1}\sum_{m^\prime = -l^\prime}^{l^\prime}
\sum_{s^\prime =\pm 1}; \nonumber \\ \sum_{\lambda''}&=&  \sum_{l'' = 0}^{n-1}\sum_{m''=-l''}^{l''}
\sum_{s'' =\pm 1};
 \end{eqnarray}
Parameters
\begin{eqnarray}
\label{7a}
  \alpha = 0.321; \ \beta = 0.412.
\end{eqnarray}

The above fully algebraic expression with nonlinear dependence on the
occupation numbers corresponds to a model of an atom in which an electron on a
layer with quantum numbers $\lambda$ moves in the field of an effective Coulomb
potential with a charge $Z_{\lambda}$. It is equal to the nucleus charge minus
the partial screening of electrons of underlying layers (the second term in
(\ref{6})) and other electrons in the same layer (the third term in (\ref{6})).
This expression is consistent with the Hohenberg-Kohn theorem
\cite{hohenberg1964inhomogeneous}, since the energy of the system depends only
on single-particle characteristics of electrons in an atom.

Let us now investigate the conditions that allow one to determine the coefficient 5/16 and the constants
$\alpha$ and $\beta$ in (\ref{7a}). The first condition arises from the asymptotic
behavior of the atomic energy $E(Z, N)$ in the limit $Z \rightarrow \infty$. In
this limit of large $Z$, all shells can be considered fully occupied, meaning that the occupation numbers $g_{nlms} = 1$, for all shells with $n \leq n_{max}$. In
this approximation we can replace the sums with integrals in the expression for
$Z_{\lambda}$ (\ref{6}) and perform the integration. We restrict ourselves to
the terms with the highest power of $n$:
\begin{eqnarray}
\label{8}
Z_{\lambda} \rightarrow  Z_n \approx Z - \nonumber \\ \int_{n_1=1}^{n-1}\int_{l_1=0}^{n_1-1}2(2l_1 + 1)\kappa_{n,l_1} dn_1dl_1 - \nonumber\\
 \frac{5}{16}\int_0^{n-1}(4l + 1)\kappa^\prime_{n,l}dl =\nonumber \\  Z - \left(\frac{2}{3} + \frac{\beta}{5}\right) n^3 .
\end{eqnarray}

Similarly, in the expression for energy (\ref{6})
\begin{eqnarray}
\label{9}
E(Z) \approx - \int_{n=1}^{n_{max}}\int_{l=0}^{n-1}(4l + 1)\frac{Z_n^2}{2n^2}dndl \approx \nonumber \\  -\int_{n=1}^{n_{max}}Z_n^2dn \approx
\left( \frac{4}{63} - \frac{4\beta}{105} + \frac{\beta^2}{175}\right)n^7_{max} + \nonumber \\ \left(-\frac{1}{3}+\frac{\beta}{10}\right)Z n^4_{max} +Z^2 n_{max}.
\end{eqnarray}

The value of $n_{max}$ is determined from the normalization condition in the same approximation:
\begin{eqnarray}
\label{10}
Z \approx \int_{n=1}^{n_{max}}2n^2dn \approx \frac{2}{3} n^3_{max}; \ n_{max}\approx \left(\frac{3Z}{2}\right)^{1/3}.
\end{eqnarray}

Then the final result is:
\begin{eqnarray}
\label{11}
E \approx - \frac{9}{28}12^{2/3}\left(1+\frac{\beta}{10}
+ \frac{\beta^2}{50}\right)Z^{7/3}.
\end{eqnarray}

If we now compare this expression with the well-known exact asymptotics of the
binding energy of a nonrelativistic atom derived from the Thomas-Fermi equation
\cite{LandauQM}.
$$
E \approx - 0.768745 Z^{7/3},
$$
we find the value $\beta = 0.412$, which is used in formula (\ref{7a}).

The physical meaning of the coefficient 5/16 and the parameter $\alpha$ describe the screening of the
nucleus charge by the electrons. Their values in the formula (\ref{7a}) are
chosen from the conditions that the largest matrix elements for transitions $(10;10)$ and $(10;20)$ calculated by (\ref{3},\ref{4}) coincide with their values calculated by using Coulomb orbitals.

Thus, the accuracy of the approximation of the original Hamiltonian by the
operator (\ref{6}) depends on the error in replacing matrix elements
(\ref{3},\ref{4}). Table \ref{table:1} presents the results of calculation of
the matrix elements using the exact formulas \cite{Skoromnik_2017} and the
approximate ones (\ref{3},\ref{4}) in the range of quantum numbers, where the values
of matrix elements can be large.

\begin{table}[h!]
\centering
\caption{Table of difference between exact \cite{Skoromnik_2017} and approximate (\ref{3},\ref{4}) matrix elements.}
\begin{tabular}{|c| c c c c c c| }
\hline
\diagbox[width=3em]{\textbf{n'l'}}{\textbf{nl}} & \textbf{10} & \textbf{20} & \textbf{21} & \textbf{30} & \textbf{31} & \textbf{32} \\ \hline
\textbf{10} & $0$ & $0$ & $0.0845$ & $0.0003$ & $0.0198$ & $0.0424$ \\
\textbf{20} & $0$ & $0.0059$ & $0.0103$ & $0.0151$ & $0.0010$ & $0.0349$ \\
\textbf{21} &$ 0.0845$ & $0.0103$ & $0.0072$ & $0.0125$ & $0.0090$ & $0.0419$ \\
\textbf{30} & $0.0003$ & $0.0151$ & $0.0125$ & $0.0030$ & $0.0038$ & $0.0059$ \\
\textbf{31} & $0.0198$ & $0.0010$ & $0.0090$ & $0.0038$ & $0.0018$ & $0.0011$ \\
\textbf{32} & $0.0424$ & $0.0349$ & $0.0419$ & $0.0059$ & $0.0011$ & $0.0036$ \\ \hline
\end{tabular}
\label{table:1}
\end{table}

It shows the difference between the two-particle matrix elements of
the interaction operator between electrons calculated in the Coulomb
orbitals basis \cite{Skoromnik_2017},
\begin{eqnarray}
\label{11a}
M_{nl}^{n^\prime l^\prime} = \frac{1}{Z}\int d\vec r_1\int d\vec r_2 \times \nonumber \\ |\psi_{nl}(\vec r_1)|^2 \frac{1}{|\vec r_1-\vec r_2 |}|\psi_{n^\prime l^\prime}(\vec r_2)|^2
\end{eqnarray}
and their approximate expressions (\ref{3},\ref{4}) through single-particle operators.

As shown by the results in this table, the average error $\delta$ for
such a substitution is $\delta < 0.01$ a.e.u. (atomic energy
units). Then, for an atom with $Z$ electrons, the contribution of the
electron-electron interaction to the total energy is determined with
an accuracy $\Delta E < \frac{1}{2}Z^2 \delta$ a.e.u. Considering that
the total energy of such an atom is approximately equal to $E \approx Z^{7/3}$
a.e.u., we find that the relative accuracy of the formula (\ref{11a})
in calculation of the atomic energy is $\Delta E /E < Z^{-1/3} \%$. This
estimate is confirmed by the numerical evaluation, which is presented in the
next section.

\section{Calculation of the atomic characteristics}
\label{sec:cal-char}
\subsection{Binding energy}

\begin{figure}[h]
	\includegraphics[width=1\linewidth]{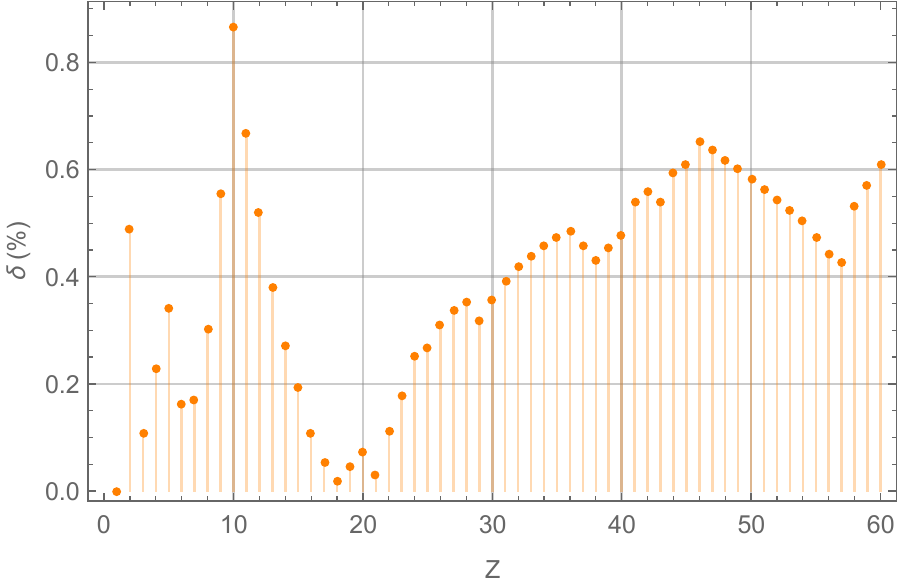}
	\caption{(Color online) Relative error (in percentages) between our results (\ref{6}) and those of the HFM.}
	\label{fig:1}
\end{figure}

Let us compare our results (\ref{6}) with the ones via HFM.  As shown in Table
\ref{tab:table2} and Fig.\ref{fig:1}, the algebraic formula (\ref{6})
approximates the results obtained by the HFM for the binding energy of atoms
with an accuracy of $\delta < 1 \%$ for all $Z$. The results are presented for
$Z\leq 60$, when the relativistic corrections are still sufficiently small. It
is important to emphasize that the experimental configurations correspond to
the minimal atomic energy calculated by the formula (\ref{6}). Additionally, it
is worth noting that our model correctly predicts the filling of electronic
shells and corresponds to the Madelung-Klechkowski rule \cite{Madelung1936,
KlechkovskiiA1962justification}

\subsection{Ionization potential}

In order to determine the applicability of our results for ions, we
calculate the values of partial ionization potentials $(IP_{nl})$ when
an electron is removed from the $(n,l)$ shell
\cite{Yeh:1985fij}. According to Eq.~(\ref{5}), we find
\begin{eqnarray}
\label{12}
IP_{n,l} = E(Z,N-1_{n,l}) - E(Z,N) .
\end{eqnarray}

\begin{figure*}[tb!]
	\includegraphics[width=0.32\linewidth]{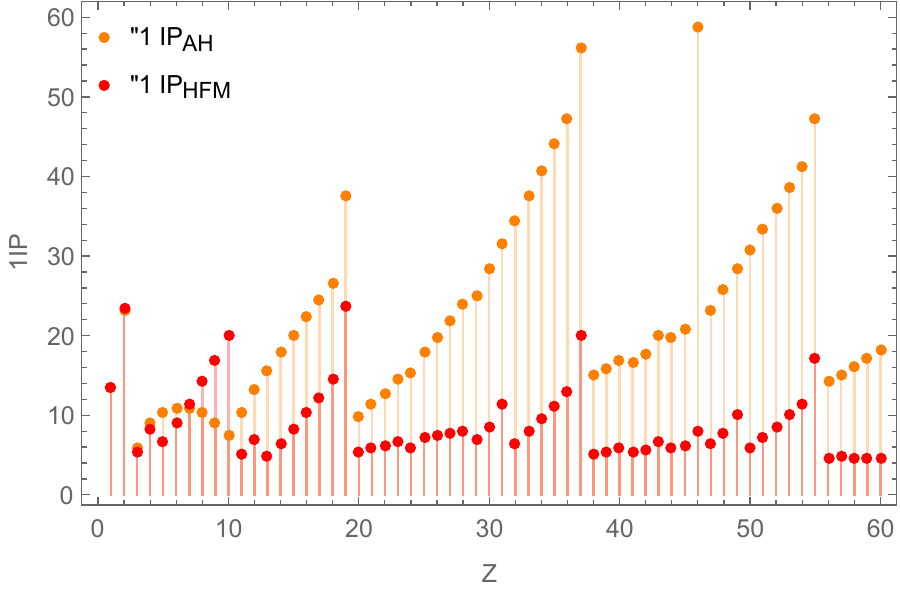}
	\includegraphics[width=0.32\linewidth]{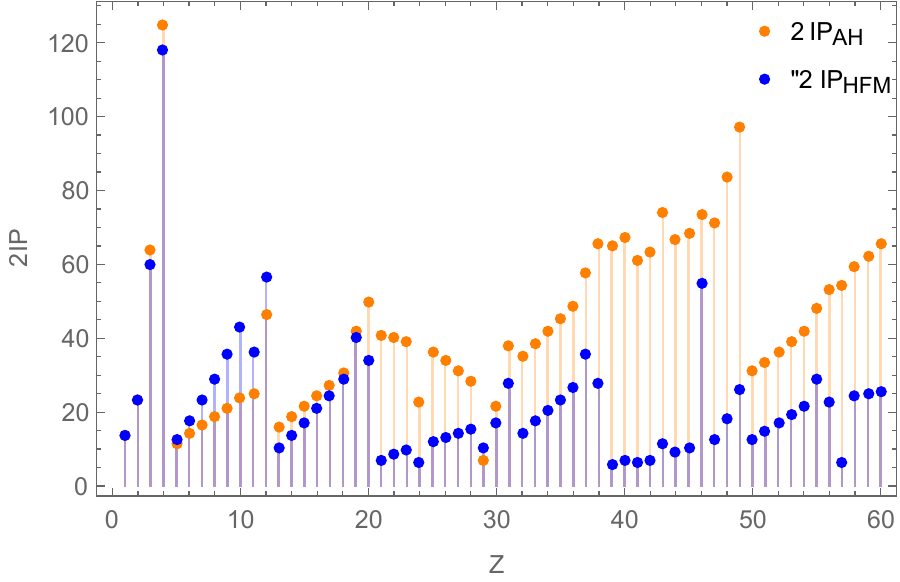}
	\includegraphics[width=0.32\linewidth]{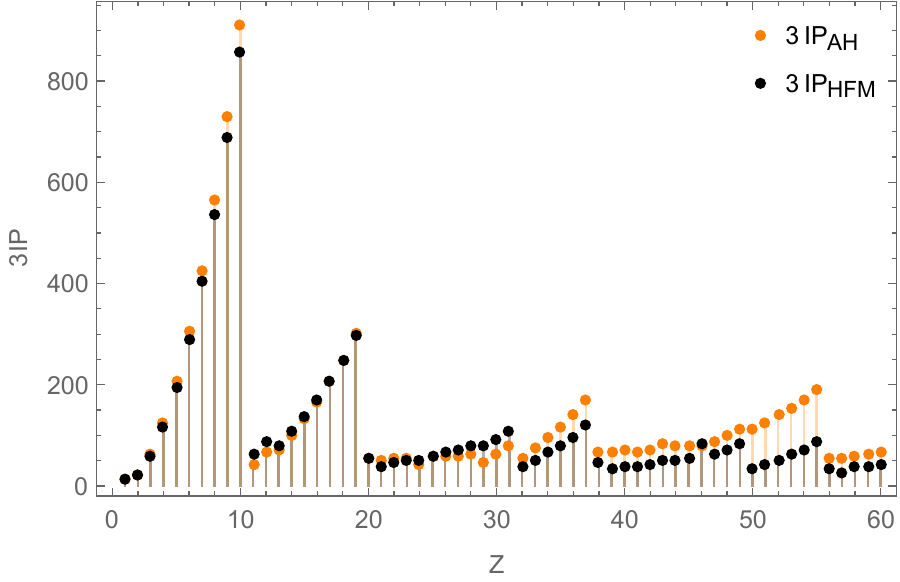}
	\caption{(Color online) Comparison of the first three ionization potentials obtained from our results and the HFM.}
	\label{fig:2}
\end{figure*}

In Table \ref{tab:table2} and Fig.\ref{fig:2}, the results for the
calculation of the first three ionization potentials (IP) are
presented. The value of the first IP is comparable in accuracy to that
of the binding energy calculations. This means that the
computed values are only in qualitative agreement with the HFM
results. However, the second and third IPs are calculated with the
relative accuracy $\delta$.

\begin{figure*}[tb!]
	\includegraphics[width=0.45\linewidth]{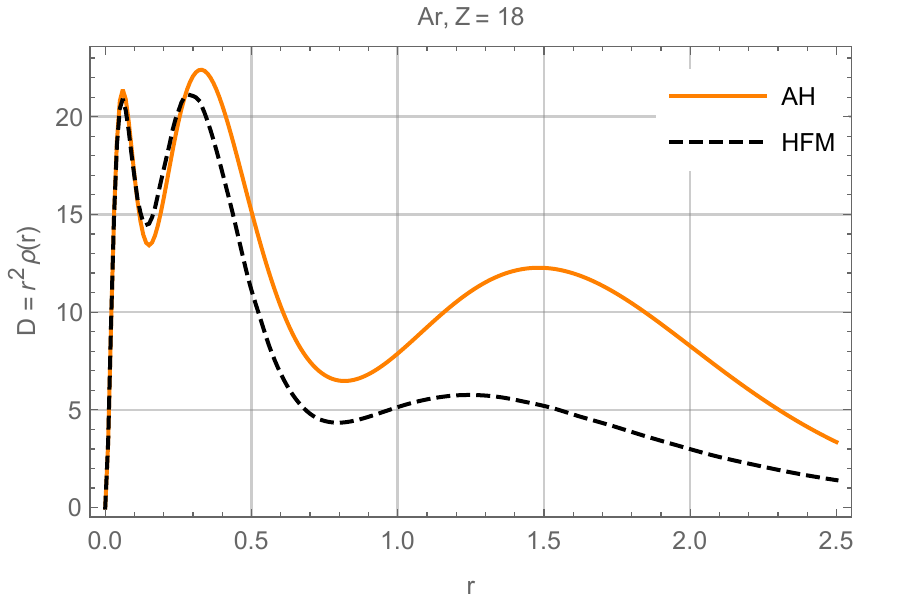}
	\includegraphics[width=0.45\linewidth]{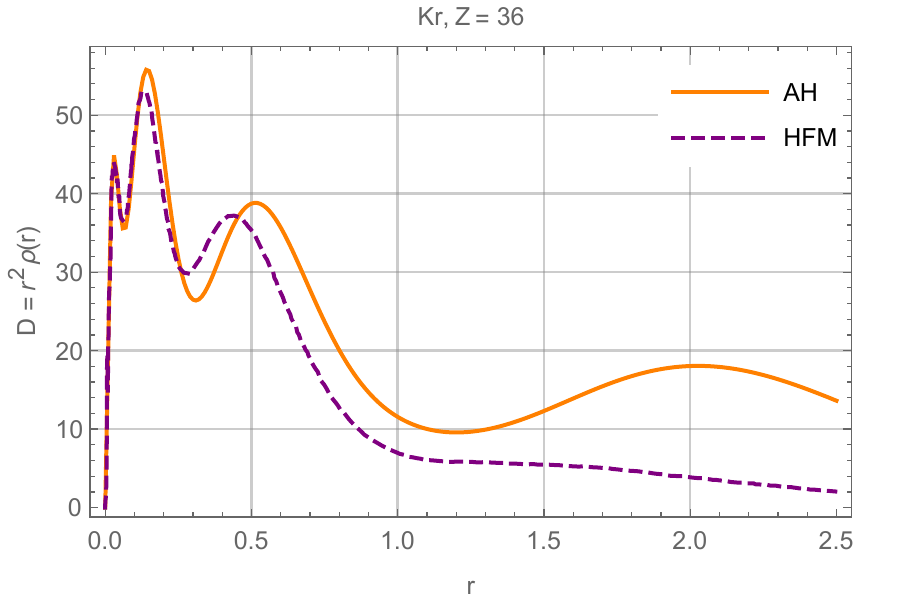}
	\caption{(Color online) The radial electron densities for argon (the left panel) and krypton (the right panel), compared with those obtained from the HFM.}
	\label{fig:3}
\end{figure*}

\begin{figure}[h]
	\includegraphics[width=1\linewidth]{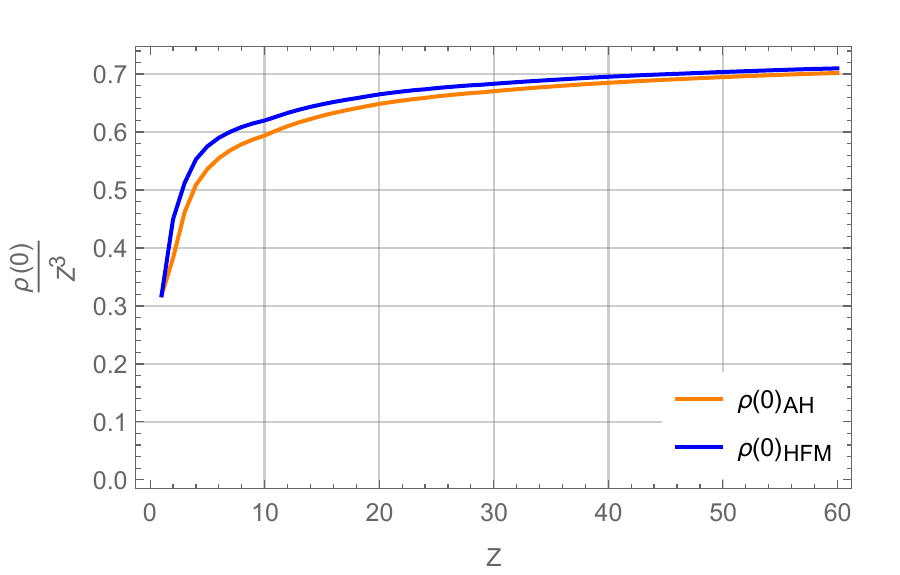}
	\caption{(Color online) Comparison of the electron density at the atomic nucleus as computed by our method and by the HFM.}
	\label{fig:4}
\end{figure}

\begin{figure*}[tb!]
	\includegraphics[width=0.45\linewidth]{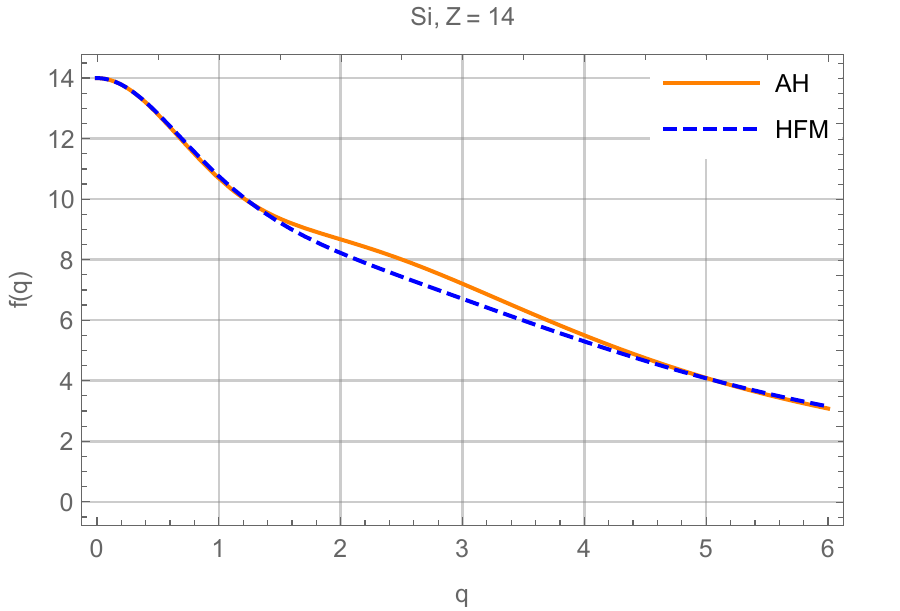}
	\includegraphics[width=0.45\linewidth]{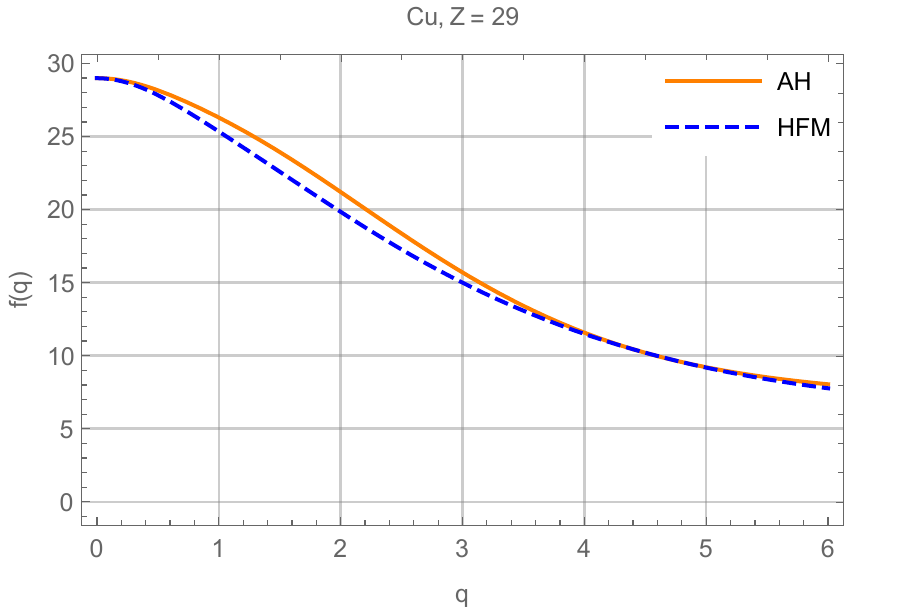}
	\caption{(Color online) The atomic scattering factors for neutral Si (the left panel) and Cu (the right panel). Orange solid and blue dashed lines indicate our results and HFM values, respectively.}
	\label{fig:5}
\end{figure*}

\subsection{Electron density distribution}

According to the definition of the density operator in the second quantization representation

\begin{eqnarray}
\label{13}
\hat \rho(\vec r) = \hat \Psi^\dagger (\vec r)\hat \Psi (\vec r); \nonumber \\ \hat \Psi (\vec r) = \sum_{\lambda}[\hat{a}_{\lambda}\psi_{\lambda}(\vec r,Z) + \hat{a}^\dagger_{\lambda}\psi^*_{\lambda}(\vec r,Z)],
\end{eqnarray}
we find its observable value in the state $\ket{\{g_{\lambda}\}}$
\begin{eqnarray}
\label{14}
  \rho(\vec r) =   \sum_{\lambda}g_{\lambda}|\psi_{\lambda}(\vec r,Z_{nl})|^2.
\end{eqnarray}
Here $Z_{nl}$ is the effective charge of the shell calculated by the formula (\ref{6}), and the normalized Coulomb orbitals are defined by the well-known formulas

\begin{eqnarray}
\psi_{\lambda}(\vec r,Z) = N_{nl}R_{nl}(r,Z)Y_{lm}(\theta, \phi)\chi_s ;\nonumber\\
R_{nl}(r)=\left(\frac{2Zr}{n}\right)^le^{-\frac{Zr}{n}} F\left(-n+l+1, 2l+2,\frac{2Zr}{n}\right), \nonumber
\end{eqnarray}
\begin{eqnarray}
\label{15}
N_{nl}=\frac{1}{(2l+1)!}\sqrt{\frac{(n+l)!}{2n(n-l-1)!}}\left(\frac{2Z}{n}\right)^{3/2}.
\end{eqnarray}
here $F(a,c,x)$ is the confluent hypergeometric
function. $Y_{lm}(\theta, \phi)$ is the spherical harmonic, $\chi_s$
is the spinor.

In particular, for atoms with a closed shell, the electron density is
spherically symmetric and is given by:
\begin{eqnarray}
\label{16}
  \rho(r) =   \sum_{n l} \frac{2l+1}{2\pi}|N_{nl}R_{nl}(r,Z_{nl})|^2.
\end{eqnarray}

Fig.\ref{fig:3} shows a comparison of the radial electron density
$D(r) = r^2 \rho (r)$ calculated based on the function (\ref{16}) and
within the framework of the HFM.

Another important observable characteristic for various applications (such as isomer shift
\cite{FILATOV2009594}, nuclear magnetic resonance
\cite{NMRPhysRevA.52.3628} ) is the electron
density at the atomic nucleus (as $r \rightarrow 0$). In our approximation, only orbitals with $l=0$ contribute to this
quantity. Thus we find the expression
\begin{eqnarray}
\label{17}
  \rho(0) =   \sum_{n00s} \frac{g_{n00s}}{4\pi}|N_{n0}R_{n0}(0,Z_{n0})|^2 = \nonumber \\  \sum_{n00s} \frac{g_{n00s}}{\pi}\left(\frac{Z_{n0}}{n}\right)^3.
\end{eqnarray}

Fig.\ref{fig:4} compares the results obtained on the basis of
(\ref{17}) and within the framework of the HFM.

\subsection{Atomic scattering factors}

The atomic scattering factors, which are of significant importance for X-ray
crystallography \cite{tablesCrystallographyC2006}, are determined by the
Fourier transform of the electron density. In our model, they can be calculated
analytically using the formulas given in \cite{Skoromnik_2017}.

\begin{eqnarray}
\label{18}
  f(q) = \sum_{nl}g_{nl} F_{nl}(Z_{nl},q); \quad \xi_{nl} = \frac{2Z_{nl}}{n},
\end{eqnarray}

\begin{eqnarray}
\label{19}
  F_{nl} =  - C_{nl}\xi^{2l+3}\frac{(n-l-1)!(n+l)!}{2n} \times \nonumber \\
 \sum_{k,m=0}^{n-l-1}\frac{\xi^{k+m}}{(2l+k+1)!(2l+m+1)!k!m!} \times\nonumber \\  \frac{d^{2l+1+k+m}}{d\xi^{2l+1+k+m}}\frac{1}{\xi^{2}+q^{2}}.
\end{eqnarray}
where $f(q)$ is the atomic scattering factor, $q$ is the transverse wave vector associated with the scattering angle $\theta$
$$
q=4\pi s a_B; \ s = \sin \theta/\lambda [\text{\AA}^{-1}],
$$
$a_B = 0.529177 \text{\AA}^{-1}$ the Bohr radius and $C_{nl}$ is the normalization coefficient
at $q = 0$
\begin{eqnarray}
\label{C_normal}
C_{nl}=\Bigg[\sum_{k,m=0}^{n-l-1}\frac{\xi^{k+m}}{(2l+k+1)!(2l+m+1)!k!m!} \times\nonumber \\ \xi^{2l+3}\frac{(n-l-1)!(n+l)!}{2n} \frac{d^{2l+1+k+m}}{d\xi^{2l+1+k+m}}\frac{1}{\xi^{2}}\Bigg ]^{-1}
\end{eqnarray}

In Fig.\ref{fig:5}, the results calculated using formula (\ref{18}) are compared with numerical data obtained from solutions of the HFM \cite{Waasmaier:sh0059}

\section{Conclusion}
\label{sec:conclusion}

In our work we developed a new approach on how to approximate the solutions of
the self-consistent Hartree-Fock equations for nonrelativistic atoms and ions.
It is based on the approximating Hamiltonian $\hat{H}_0 $, whose eigenfunctions
are given by the Coulomb orbitals and are a good approximation to HFM
solutions.  The Hamiltonian is represented by a sum of single-particle
operators, and its eigenvalues are determined by a fully algebraic expression.
They offer a good approximation for the atomic binding energy. It is shown that
within our approach one can analytically calculate various observable
characteristics of atoms and ions. In particular, ionization potential,
electron density distribution, and atomic scattering factor are in good
agreement with the results obtained in the HFM approximation.

Furthermore, it is important to emphasize that the representation of the
single-electron approximation in the form of an eigenvalue problem instead of a
system of nonlinear integro-differential equations allows one to use the
standard form of quantum mechanical perturbation theory over the operator
$\hat{V}=\hat{H}-\hat{H}_0$. This perturbation operator takes into account
correlation effects in multi-electron systems. We also suppose that our
approach can be generalized for molecules.


\section{Acknowledgments}

The authors are grateful to Prof. A.P.Ulyanenkov  for the support of this work.

This work is funded by the Hue University under grant no. DHH2025-18-05. We also acknowledge the generous support of Vingroup JSC and the Postdoctoral Scholarship Program of the Vingroup Innovation Foundation (VINIF), Institute of Big Data which provided funding for Nguyen Quang San.

\begingroup
\squeezetable
\begin{table*}
\caption {\label{tab:table2} Calculated binding energy and the first three ionization potentials, compared with values obtained from the HFM.}
\begin{ruledtabular}
\begin{tabular}{llllllllllll}
 $Z$ & $E_{\text{AH}}$ & $E_{\text{HF}}$ & $IP_{\text{AH}}$ & $IP_{\text{HF}}$ & Config. &  $Z$ & $E_{\text{AH}}$ & $E_{\text{HF}}$ & $IP_{\text{AH}}$ & $IP_{\text{HF}}$ & Config.                                                \\    \hline
1& 0.5 & 0.5 &13.6 & 13.6 &$1s$ & 2& 2.84766 &2.86168 &23.0563& 23.4& $1s$ \\
3& 7.44078 &7.43273& 64.1190 & 59.8 & $1s$ & 3 & 7.44078 &7.43273& 5.93303 & 5.5 & $2s$ \\
4& 14.6062 & 14.5730& 124.532 & 118.3 & $1s$ & 4 & 14.6062 &14.5730& 9.11561 & 8.2 & $2s$ \\  \arrayrulecolor[gray]{0.65}\hline
& & & 205.032 & 195.5& $1s$ & & & & 305.618 & 290.9 &$1s$ \\
5& 24.6124 & 24.5291 & 11.5599 & 12.6 & $2s$ & 6 &  37.7498 & 37.6886 & 13.9960 & 17.5 & $2s$ \\
& &   & 10.2556 & 6.7 & $ 2p$ & &  & & 10.8129 & 9 & $2p$ \\  \hline
& & & 426.291 & 404.6 & $1s$ &  & & & 567.050 & 536.8 &$1s$ \\
7 & 54.3092 & 54.4009 & 16.4238 & 23.1 & $2s$& 8 & 74.5820 & 74.8094 & 18.8434 & 29.2 & $2s$ \\
 &  &  & 10.8159 & 11.5 & $2p$ & & & & 10.2646 & 14.2 & $2p$  \\ \hline
& & & 727.896 & 687.6 & $1s$ &  & & & 908.828 & 857 & $1s$ \\
9 & 98.8592 & 99.4093 & 21.2547 & 35.8 & $2s$ & 10 & 127.432 & 128.547 & 23.6577 & 43.1 & $2s$ \\
 &  &  & 9.15886 & 17 & $2p$ & & & & 7.49878 & 20 & $2p$ \\ \hline
& & & 42.9851 & 64.3 & $2s$& & & & 65.4015 & 89.1& $2s$ \\
11 & 160.779 & 161.859 & 25.1571 & 36.3 & $2p$ & 12 & 198.578 & 199.615 & 46.2854 & 56.4 & $2p$ \\
& & & 10.3722  & 5.1 & $3s$ & & & & 13.2589 & 6.9 & $3s$ \\ \hline
& & & 71.0224 & 80.9 & $2p$ & & & & 99.3681 & 108.2 & $2p$ \\
13& 240.962 & 241.877 & 15.9578 & 10.1 & $3s$ & 14 & 288.076 & 288.854 & 18.7241 & 13.6 & $3s$ \\
& & & 15.5861 & 4.9 & $3p$& & & & 17.8643 & 6.5& $3p$ \\ \hline
& & & 131.322 & 138.5 & $2p$ & & & &166.886 & 171.8 & $2p$\\
15 & 340.065 & 340.719 & 21.5579 & 17.1& $3s$ & 16 & 397.072 & 397.505 & 24.4590 & 20.8 & $3s$ \\
& & & 20.0958 & 8.3 & $3p$ & & & & 22.2806 & 10.3 & $3p$ \\ \hline
& & & 206.057 & 208.2 & $2p$ & & & & 248.838 & 247.7 & $2p$ \\
17 & 459.243 & 459.482 & 27.4276 & 24.7 & $ 3s $ & 18 & 526.722 & 526.818 & 30.4636 & 28.7 & $3s$ \\
& & & 24.4188 & 12.3 & $3p$ & & & & 26.5103 & 14.5 & $3p$ \\ \hline
& & & 303.675 & 299.4 & $2p$ & & & & 54.0049 & 52.7& $3s$ \\
19 & 598.898 & 599.165 & 41.7981 & 40.2 & $3s$ & 20 & 676.256 & 676.758 & 49.8269 & 33.8 & $3p$ \\
& & & 37.6722 & 23.6 & $3p$ & & & &9.88174 & 5.4 & $4s$ \\ \hline
& & & 51.7948 & 39.2 & $3p$ & & & & 53.6158 & 44.6 & $3p$\\
21& 759.957 & 759.736 & 11.3018 & 5.9 & $4s$& 22 & 849.356 & 848.406 & 12.8156 & 6.2& $4s$ \\
& & & 41.0530 & 7.2 & $3d$ & & & & 40.3368 & 8.5 & $3d$\\ \hline
& & & 55.2901 & 49.9 & $3p$ & & & & 43.8421 & 50.2 & $3p$\\
23& 944.575 & 942.884 & 14.4232 & 6.6 & $4s$& 24 & 1045.97 & 1043.36 & 15.2649 & 5.9 & $4s$ \\
& & & 39.2656 & 9.8 & $3d$ & & & & 22.4789 & 6.5 & $3d$ \\ \hline
& & & 58.1985 & 60.9 & $3p$ & & & & 59.4326 & 66.6 & $3p$\\
25 & 1152.95 & 1149.87 & 17.9200 & 7.1 & $4s$ & 26 & 1266.35 & 1262.44 & 19.8091 & 7.4 & $4s$ \\
& & & 36.0579 & 12 & $3d$ & & & & 33.9215 & 13.1 & $3d$ \\ \hline
& & & 60.5199 & 72.4 & $3p$ & & & & 61.4604 & 78.3 & $3p$\\
27& 1386.06 & 1381.41 & 21.7921 & 7.7 & $4s$ & 28 & 1512.18 & 1506.87 & 23.8688 & 7.9 & $4s$ \\
& & & 31.4300 & 14.2 & $3d$ & & & & 28.5834 &15.2 & $3d$ \\ \hline
& & & 46.3432 & 77.7 & $3p$ & & & & 62.9013 & 90.6 & $3p$ \\
29 & 1644.14 & 1638.96 & 24.8999 & 6.9 & $4s$ & 30 & 1784.18 & 1777.85 & 28.3038 & 8.4 & $4s$ \\
& & & 6.70433 & 10.1 & $3d$ & & & & 21.8252 & 17.1 & $3d$ \\ \hline
& & & 80.4997 & 107.4 & $3p$ & & & & 34.9564 & 14.4 & $4s$ \\
31& 1930.79 & 1923.26 & 31.6036 & 11.4 & $4s$ & 32 & 2084.06 & 2075.36 & 55.8941 & 39.3 & $3d$ \\
& & & 38.2218 & 27.8 & $3d$ & & & & 34.5307 & 6.4 & $4p$ \\ \hline
& & & 38.3621 & 17.4 & $4s$ & & & & 41.8208 & 20.3 & $4s$\\
33& 2244.06 & 2234.24 & 74.8423 & 52 & $3d$ & 34 & 2410.89 & 2399.87 & 95.0662 & 65.8 & $3d$ \\
& & & 37.6682 & 7.9 & $4p$& & & &40.8218& 9.5& $4p$ \\ \hline
& & & 45.3324 & 23.4& $4s$& & & & 48.8970 & 26.5& $4s$ \\
35 & 2584.63 & 2572.44& 116.566 & 80.6& $3d$ & 36 & 2765.36 & 2752.05 & 139.341 & 96.6 & $3d$ \\
& & & 43.9916 & 11.2& $4p$ & & & & 47.1776 & 13 & $4p$ \\ \hline
& & & 57.7859 & 35.5 & $4s$ & & & & 66.9749 & 45& $4s$ \\
37 & 2951.79 & 2938.36 &  169.701 & 120.7 & $3d$ & 38 & 3145.01 & 3131.55 & 65.3937 & 27.7 & $4p$\\
& & & 56.1109 & 20.1 & $4p$ & & & & 15.0022 & 5 & $5s$ \\ \hline
& & & 68.8655 & 32.6 & $4p$ & & & & 72.3177 & 37.1 & $4p$ \\
39 & 3346.83 & 3331.68 & 15.9524 & 5.5 & $5s$ & 40 & 3555.90 & 3539 & 16.9316 & 5.9& $5s$ \\
& & & 64.7830 & 5.6 & $4d$ & & & & 67.2576 & 7 & $4d$ \\ \hline
& & & 68.4099 & 38.5 & $4p$ & & & & 71.4355 & 42.5 & $4p$ \\
41 & 3773.90 & 3753.6 & 16.7431 & 5.5 & $5s$ & 42 & 3997.75 & 3975.55 & 17.7426 & 5.7 & $5s$\\
& & & 61.1727 & 6.1 & $4d$ & & & & 63.0537 & 7.2 & $4d$ \\ \hline
& & & 82.5565 & 50.3 & $4p$ & & & & 77.4280 & 50.6 & $4p$ \\
43 & 4227.44 & 4204.79 & 20.0429 & 6.8 & $5s$ & 44 & 4467.97 & 4441.54 & 19.8285 & 6& $5s$\\
& & & 74.1433 & 11.2 & $4d$ & & & & 66.5468 & 9.3 & $4d$\\ \hline
& & & 80.3948 & 54.7 & $4p$ & & & & 81.0555 & 84.2 & $4s$\\
45 & 4714.49 & 4685.88 & 20.9149 & 6.2 & $5s$ & 46 & 4970.08 & 4937.92 & 73.6597 & 54.6 & $4p$ \\
& & & 68.1589 & 10.4 & $4d$& & & & 58.8496 & 8 & $4d$ \\ \hline
& & & 86.2695 & 62.9 & $4p$ & & & & 99.2287 & 72.3 & $4p$ \\
47 & 5230.75 & 5197.7 & 71.1140 & 12.6 & $4d$ & 48 & 5498.92 & 5465.13 & 25.8079 & 7.7 & $5s$ \\
& & & 23.1747 & 6.4 & $5s$ & & & & 83.8265 & 18.3 & $4d$ \\ \hline
& & & 112.558 & 84.1 & $4p$ & & & & 31.0362 & 12.5 & $5s$ \\
49 & 5774.64 & 5740.17 & 28.4028 & 10.1 & $5s$ & 50 & 6057.98 & 6022.93 & 110.652 & 34.4 & $4d$ \\
& & & 97.0060 & 26.2 & $4d$ & & & & 30.8367 & 5.9 & $5p$ \\ \hline
& & & 33.7080 & 14.8 & $5s$ & & & & 36.4182 & 17.1 & $5s$ \\
51 & 6348.97 & 6313.49 & 124.766 & 42.9 & $4d$ & 52 & 6647.68 & 6611.8 & 139.346 & 52 & $4d$ \\
& & & 33.3858 & 7.3 & $5p$ & & & & 35.9583 & 8.6 & $5p$ \\ \hline
& & & 39.1669 & 19.4 & $5s$ & & & & 41.9541 & 21.8 & $5s$ \\
53& 6954.17 & 6917.98 & 154.393 & 61.5 & $4d$ & 54 & 7268.49 & 7232.1 & 169.907 & 71.5 & $4d$\\
& & & 38.5540 & 10 &  $5p$ & & & & 41.1729 & 11.4 & $5p$\\ \hline
& & & 47.8920 & 28.8 & $5s$ & & & & 53.9346 & 35.9 & $5s$ \\
55 & 7589.54 & 7553.9 & 190.664 & 87.6 & $4d$ & 56 & 7918.28 & 7883.5 & 53.2763 &  22.9 & $5p$ \\
& & &  47.1604 & 17.1 & $5p$ & & & & 14.3954 & 4.5 & $6s$ \\ \hline
& & & 56.2566 & 26.4 & $5p$ & & & & 59.8648 & 38.4 & $5s$ \\
57& 8256.15 & 8221.1 & 14.9726 & 4.9 & $6s$ & 58 & 8612.34 & 8566.9 & 59.1845 & 24.3 & $5p$ \\
& & & 54.5356 & 6.2 & $5d$ & & & & 16.2366 & 4.6 & $6s$ \\ \hline
& & & 62.9421 & 39.6 & $5s$ & & & & 66.0942 & 40.7& $5s$ \\
59& 8972.11 & 8921.2 & 62.2508 & 25 & $5p$ & 60 & 9340.45 & 9283.9 & 65.3918 & 25.6 & $5p$ \\
& & & 17.1986 & 4.6 & $6s$ & & & & 18.1881 & 4.7 & $6s$ \\ \arrayrulecolor{black}
\end{tabular}
\end{ruledtabular}
\end{table*}
\endgroup

\newpage

\bibliography{biblapprHforHFsol}

\end{document}